# A broadband silicon photonic-integrated-circuit based RF spectrum analyzer with 10 MHz spectral resolution

Brandon Redding[1,*], Joseph B. Murray[1], Matthew J. Murray[1], Ross T. Schermer[1], Nicholas Cox[1], Sean Pang[2], Kate Musick[3], Christopher Long[3], Jayson Briscoe[4], Lewis G. Carpenter[4], Saaketh Desai[5], Nick Karl[3], Raktim Sarma[3,5,*]

[1]U.S. Naval Research Laboratory, Washington, DC, USA
[2]CREOL, The College of Optics and Photonics, University of Central Florida, Orlando, Florida, USA
[3]Sandia National Laboratories, Albuquerque, New Mexico, USA
[4]AIM Photonics, Albany, New York, USA
[5]Center for Integrated Nanotechnologies, Sandia National Laboratories, Albuquerque, New Mexico, USA
*brandon.f.redding.civ@us.navy.mil; rsarma@sandia.gov

**Abstract**

Designing miniaturized optical spectrometers is an increasingly active area of research as spectrometers are crucial components for a wide range of applications including chemical and material analysis, medical diagnostics, classical and quantum sensing, characterization of light sources, and radio frequency (RF) spectrum analysis. Among these applications, designing on-chip spectrometers for RF spectrum analysis is particularly challenging since it requires combining high resolution and large bandwidth with a fast update rate. Existing chip-scale spectrometers cannot achieve the resolution required for RF analysis, setting aside challenges in maintaining a fast update rate and broad bandwidth. In this work, we address these challenges by introducing a silicon photonic integrated circuit (PIC)-based RF spectrum analyzer that combines an ultra-high-resolution speckle spectrometer with an interferometric RF-to-optical encoding scheme. The PIC-based speckle spectrometer uses a path-mismatched multimode interferometer with inverse designed splitters to compensate for waveguide loss, enabling a record-high resolution of 100 MHz (0.8 pm at a wavelength of 1550 nm). To further improve the resolution of the overall RF spectrum analyzer, we modify the RF-to-optical encoding scheme by directing the RF signal through a path mismatched interferometer and encoding the outputs of the RF interferometer on separate optical carriers. This further reduces the RF spectral correlation width of the combined system, enabling the RF spectrum analyzer to resolve RF tones separated by 10 MHz across a bandwidth of 10 GHz. Since this approach operates as a single-shot spectrometer, it can support fast update rates, providing a path to compact, persistent wideband RF spectrum analysis.

**Introduction**

The demand for optical spectrometers with small size, weight, power, and cost (SWAP-C) is rapidly increasing[1] as spectrometers are core components in a host of emerging applications including portable classical and quantum sensing[2], hyperspectral imaging[3], lab-on-a-chip systems[4], quantum and classical optical network monitoring[5,6], and RF spectrum analysis[7–9]. While these applications have varying requirements, designing on-chip spectrometers for RF surveillance is among the most challenging since it requires combining high resolution and large bandwidth with a fast update rate and persistent wideband coverage which precludes spectrum-scanning techniques. The demand for persistent wideband RF surveillance is also rapidly increasing due to the widespread adoption of wireless communications, RADAR, and electronic warfare technologies[10,11]. Unfortunately, traditional electronic approaches for RF surveillance are limited by the ~1 GHz bandwidth of analog-to-digital converters (ADCs) capable of continuous, real-time operation[12,13]. While a bank of ADCs could be used to monitor a wider bandwidth, this brute force approach introduces unacceptable SWAP-C for most applications.

A variety of approaches have been proposed to enable persistent wideband RF spectral analysis, but none of these solutions provide the resolution (<100 MHz), bandwidth (>10 GHz), and update rate (>100 kHz) required for electronic support and RF spectral analysis applications, while maintaining an acceptably low SWAP-C (e.g., in a drone or satellite compatible form factor[14]). For example, Nyquist folding receivers[15–17] and compressive sensing schemes using pseudo random bit sequences[18,19] enable the detection of a wide bandwidth using a single ADC, but require precision, high-bandwidth electronics that drive the SWAP-C. Photonics has the potential to enable unique solutions not readily implemented in the RF domain. For example, photonic techniques for RF spectral analysis have been proposed based on spectral hole burning[20–24], frequency-to-time mapping, frequency-to-power mapping, four-wave mixing, and stimulated Brillouin scattering[25]. However, so far, these techniques lack the required performance (resolution, bandwidth, update rate, probability of detection) or rely on bulky off-chip components or cryogenic cooling which increases the SWAP-C[26].

In principle, a photonic RF spectrum analyzer could simply encode the RF signal on an optical carrier using an electro-optic modulator and then measure the RF spectrum with an optical spectrometer. The challenge with this approach is designing a low SWAP-C optical spectrometer capable of achieving the required resolution, bandwidth, and update rate. Even traditional grating-based spectrometers, despite their large size, rarely reach the target resolution of <100 MHz (<0.8 pm at a wavelength of 1550 nm) and the highest resolution on-chip spectrometer reported to our knowledge used high-Q silicon photonic racetrack resonators as tunable, narrow-band filters to achieve 5 pm resolution (625 MHz at 1550 nm)[27], which is still insufficient for many RF applications.

These limitations recently led researchers to investigate the use of novel spectrometer designs with the potential for higher resolution in a compact footprint. Speckle spectrometers[28], which rely on a dispersive medium to generate wavelength dependent speckle patterns that can be used to infer the optical spectrum, are particularly promising for this application. A 100 m multimode-fiber-based speckle spectrometer was used to construct an RF spectrum analyzer with ~100 MHz resolution[8]. Similarly, a Rayleigh-backscattering based speckle spectrometer relying on a 100 m single mode fiber achieved 25 MHz resolution across 15 GHz with a >100 kHz update rate[9]. However, both approaches remained relatively bulky due to the long fiber lengths and required frequent recalibration due to the environmental sensitivity of the fiber. For many applications, a photonic integrated circuit (PIC) based approach would be preferable, both from a SWAP perspective and because a monolithic PIC-based spectrometer could reduce the need for frequent recalibration. These potential advantages motivated researchers to develop an RF spectrum analyzer using a PIC-based speckle spectrometer constructed using a multimode waveguide instead of a multimode fiber[7]. Unfortunately, waveguide propagation loss limited the length of the waveguide to 10 cm, resulting in a spectral correlation width of several hundred MHz—too high for many RF applications.

Improving the resolution of a PIC-based spectrometer to the <100 MHz level required for RF spectral analysis is challenging for two reasons: (1) the limited physical size of a PIC and (2) the waveguide propagation loss. In general, both of these factors can limit the spread in optical pathlengths which dictates the resolution of the spectrometer[29]. To overcome the first limitation due to the PIC size, researchers have introduced spectrometers based on multiple scattering designs[30] or complex cavity structures[31–33] that allow the light to travel over a much larger distance than the physical size of the waveguiding structure. In theory, these approaches can support very long pathlengths as light circulates in a cavity or undergoes multiple scattering. However, propagation loss attenuates these longer paths, limiting the effective spread in pathlengths and therefore limiting the resolution[34]. To achieve higher resolution, we need to either reduce the propagation loss, or design a spectrometer that is less sensitive to the attenuation of these longer pathlengths. While a great deal of effort has gone into minimizing propagation loss, very few works have investigated the second approach of designing a spectrometer that preferentially couples more light into longer paths to compensate for the higher loss these paths will experience.

In this work, we introduce a PIC-based RF spectrum analyzer that combines a novel speckle spectrometer design with a modified RF-to-optical encoding scheme. The speckle spectrometer consists of a pair of path-mismatched, multimode interferometers constructed with inverse designed splitters that direct more optical power into the longer paths to compensate for propagation loss. This design allows us to trade-off optical attenuation for higher spectral resolution—providing a degree of freedom that is unavailable in existing spectrometer designs. Using standard, single-sideband RF encoding[9], our PIC-based spectrometer can resolve two RF tones separated by 100 MHz. To further improve the resolution, we modified the RF-to-optical encoding scheme by coupling the RF signal through a path-mismatched RF interferometer. The two output ports of the RF interferometer are encoded on separate optical carriers, resulting in a speckle pattern that changes rapidly with RF frequency. This approach further reduces the RF spectral correlation width of the combined system and enables the RF spectrum analyzer to resolve two lines separated by only 10 MHz. The RF spectrum analyzer was calibrated and tested in 5 MHz steps across a 10 GHz band from 10 to 20 GHz (the bandwidth was limited by available lab equipment, not the spectrometer design). By combining a record high-resolution PIC-based speckle spectrometer design with a novel RF-to-optical encoding scheme, this work provides a path to low-SWAP-C RF spectrum analysis with the required resolution, bandwidth and update rate for demanding RF surveillance and electronic support applications.

**Results**

We designed the PIC-based speckle spectrometer to meet the bandwidth (10+ GHz), update rate (100+ kHz), and spectral resolution (<100 MHz) required for RF spectral analysis applications. The bandwidth of a speckle spectrometer is dictated by the number of measured speckle grains, the spectral correlation width, and the sparsity of the signal being measured, assuming a compressive sensing based spectral reconstruction[35]. To cover 10 GHz bandwidth with 10 MHz resolution, we would need to recover a spectrum with 1000 spectral channels, which is relatively modest considering that state-of-the-art speckle spectrometers can reconstruct >10,000 spectral channels[36]. Assuming we can tolerate a compression ratio of 10 (a typical value used in speckle spectrometers[9,31]), we would need to detect 100 speckle grains to reconstruct 1000 spectral channels. In our design, this implies that the multimode waveguides producing the speckle grains should support 100 modes which is easily achieved. Current photonic foundry processes can also support the integration of ~100 photodetectors on a PIC[37,38]. Thus, achieving the required bandwidth for RF spectral analysis is not a major hurdle. The update rate of a PIC-based speckle spectrometer is dictated by the bandwidth of the photodetectors (assuming single-shot operation[31] as opposed to temporally modulated speckle spectrometers[32,36]). Since integrated detectors on a PIC can have bandwidth >>100 kHz[37], the update rate is also easily satisfied. However, achieving spectral resolution below 100 MHz on a PIC is much more challenging.

**Optimizing the Resolution of a PIC-based Speckle Spectrometer**

While a variety of PIC-based spectrometers designs have been proposed, their resolution is eventually limited by waveguide propagation loss. In this work, we introduce a design which provides the ability to trade-off optical insertion loss for resolution on a PIC with fixed propagation loss. To illustrate the advantage of this scheme, we used a simple model to compare the spectral correlation width of a speckle spectrometer based on disordered cavity type designs[31–33] with the path-mismatched, multimode interferometer design presented in this work.

The resolution of cavity-based spectrometers is not limited by the physical size of the PIC, since light can recirculate in the same structure. Instead, optical attenuation limits the effective pathlength of light in the cavity, and thus limits the spectral correlation width which limits the spectral resolution[33]. To illustrate this effect, we modeled the spectral correlation width for a disordered cavity-based speckle spectrometer as a function of propagation loss. In our simplified model, light is coupled into a disordered structure supporting many potential optical paths between the input and the detector, as shown in Fig. 1a. This model is useful for understanding a variety of cavity-based speckle spectrometers including spectrometers based on a arrays of coupled micro-rings[33], coupled microdisk resonators[32], or evanescently coupled multimode spirals[31]. In each case, we can model the measured speckle pattern as a summation of the electric field contribution from many paths as

$$E_{det}(f_{opt}, m) = \sum_{s=1}^{N_s} A_{s,m} e^{-(\alpha/2)L_{s,m}} e^{i[2\pi L_{s,m} f_{opt}/(c/n)]}$$

where $E_{det}$ is the total electric field reaching the $m^{th}$ detector and is calculated as a function of optical frequency, $f_{opt}$. The field includes contributions from $N_s$ paths of length $L_{s,m}$ and amplitude $A_{s,m}$, with $\alpha$ the attenuation coefficient, $c$ the speed of light, and $n$ the effective refractive index. The measured speckle intensity can be expressed as $I_{det}(f_{opt}, m) = |E_{det}(f_{opt,m})|^2$. For each detected spatial or temporal speckle grain $m$, the light reaching the detector is a summation of light following $N_s$ scattering paths. Each path has a randomly assigned amplitude $A_{s,m}$ between 0 and 1 and a random path length, $L_{s,m}$ between 0 and $L_{max}$. The electric field amplitude reaching the detector from each path is attenuated by $e^{-(\alpha/2)L_{s,m}}$ (the detected optical power contribution from each path is attenuated as $e^{-\alpha L_{s,m}}$). The last term in the summation accounts for the optical phase accumulated along the path, which is optical frequency dependent. In our model, we set $N_s$ to be $10^3$ and $L_{max}$ to be $L_{max} = -\ln[\alpha_{max}]/\alpha$, where $\alpha_{max} = 10^{-4}$ is the loss for the longest simulated path length (any pathlengths beyond this have a negligible contribution on the spectral correlation width of the measured speckle pattern). We assumed an effective index of $n = 3$.

Using this model, we simulated the speckle patterns produced as a function of optical frequency for varying attenuation coefficient, $\alpha$. The spectral correlation function[30] for a disordered cavity with a propagation loss of 1 dB/cm is shown in Fig. 1c. In this case, the correlation function has a half-width at half-maximum (HWHM) of ~350 MHz—too high for most RF spectrum analysis applications. Improving the resolution is possible if the propagation loss, $\alpha$, can be reduced. In Fig. 1c, we present the HWHM correlation width as a function of propagation loss, showing that a 10x reduction in loss could enable a 10x improvement in resolution. The limitation of this disordered cavity approach is that longer optical pathlengths always have lower amplitude than the shorter pathlengths. As a result, the spectral correlation width is set by the propagation loss and lower loss waveguides are required to improve the spectrometer resolution.

In addition, this simulation represents the best-case performance for this type of disordered or chaotic cavity type speckle spectrometer. In particular, it assumes that the input and output waveguides do not introduce any additional loss (weak perturbative regime), and that the circulating modes only experience the attenuation coefficient, $\alpha$. In the chaotic cavity type schematic shown in Fig. 1a, this might be accomplished by using a very large cavity relative to the size of the input/output waveguide so that loss from the waveguides has a negligible effect on the cavity lifetime. Using evanescently coupled output waveguides to couple light in and out of the cavity could also approach this limit if the coupling gap is large enough not to perturb the cavity Q (this trend was investigated in Ref. [32]). In practice, both these cases come with a trade-off since less light will reach the detectors as the design approaches the spectral correlation limit. As a result, cavity type speckle spectrometers are unlikely to reach the limit shown in Fig. 1d, and indeed, the highest-resolution cavity based speckle spectrometers reported to date have been limited to a spectral correlation width of ~1 GHz [31,32].

In this work, we introduce a path-mismatched interferometer scheme that allows us to explicitly direct more power into the longer path-lengths to compensate for propagation loss, as shown in Fig. 1b. This approach circumvents the challenge of coupling light in and out of a cavity without reducing the cavity lifetime. We can use the same basic model to analyze the expected spectral correlation width using this scheme (see Methods for details). The advantage of this approach is that we can adjust the coupling coefficients into the two optical paths, $\kappa_{short}$ and $\kappa_{long}$ (where $\kappa_{short} + \kappa_{long} = 1$). Here, we set the coupling coefficients to balance the power after transmission through the two arms as $\kappa_{short} = 1/[e^{\alpha \Delta L} + 1]$. Assuming the short path length is negligible so that $\Delta L \approx L_{s-long}$, the average loss for light taking the longer path will be $\alpha_{max} = \alpha \Delta L$. In this case, we can choose the path mismatch based on the maximum optical loss we are willing to accept. This provides a degree of freedom that is unavailable in the disordered cavity design. For example, if we can tolerate 20 dB of loss based on the expected input power and acceptable SNR, we can use an inverse-designed splitter to direct 99% of the input power into the long arm of a path-mismatched interferometer so that the power between the long and short arm are balanced when they are recombined at the output. This is a particularly attractive trade-off in the RF spectral analysis application targeted here, in which the seed laser power can be increased to compensate for waveguide loss.

We simulated the path-mismatched interferometer scheme for $\alpha_{max} = 0.25, 0.1, or\ 0.01$ to illustrate the effect of changing the allowed attenuation through the long pathlength. An example spectral correlation function is shown in Fig. 1c for $\alpha_{max} = 0.1$ and attenuation coefficient $\alpha = 1$ dB/cm. Compared to the idealized disordered cavity with the same 1 dB/cm loss, the path-mismatched cavity enables a ~2x reduction in the spectral correlation width. As shown in Fig. 1d, this advantage holds as a function of $\alpha$, provided the path-mismatch is adjusted to maintain a fixed $\alpha_{max}$. Increasing the acceptable loss to $\alpha_{max} = 0.01$ enables a ~4x reduction in the spectral correlation width compared to the idealized disordered cavity with the same propagation loss. This simple model shows the potential for a path-mismatched speckle spectrometer to overcome the resolution limit for a disordered cavity-based spectrometer using a fabrication process with the same attenuation coefficient.

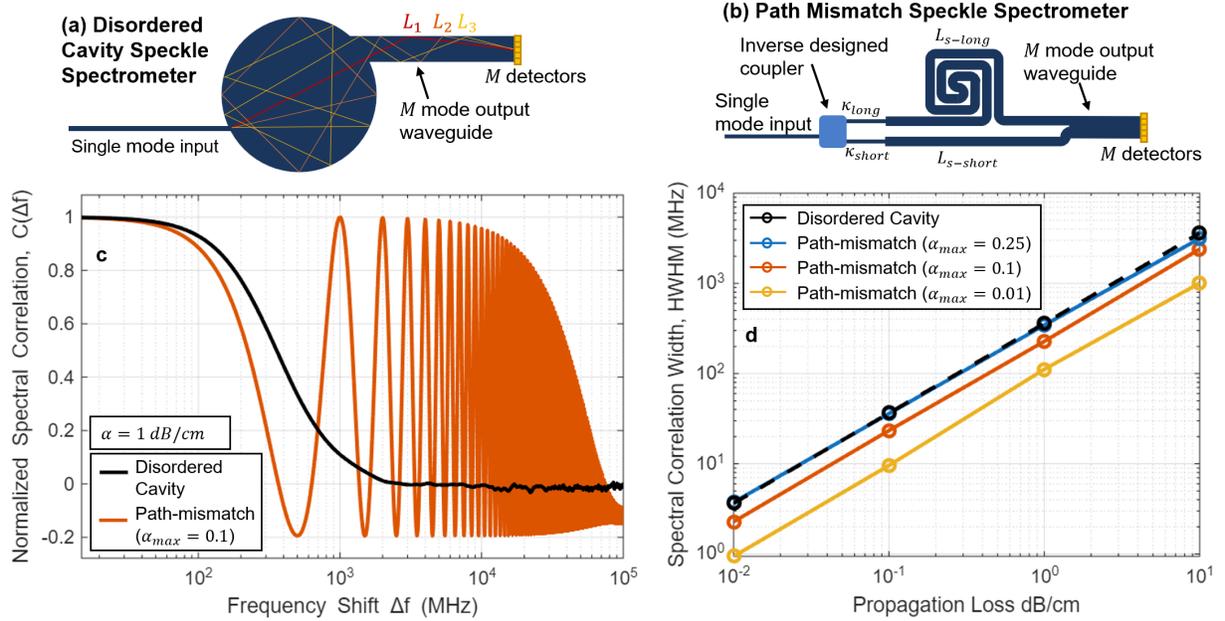

**Fig. 1. Simulated spectral correlation width vs. speckle spectrometer design. a,b** Schematic showing the layouts of speckle spectrometers based on a disordered cavity and path mismatched interferometer. **c** Comparison of calculated spectral correlation functions of the disordered cavity and path mismatch spectrometers for $\alpha_{max} = 0.1$ and attenuation coefficient $\alpha = 1$ dB/cm. **d** HWHM extracted from the calculated spectral correlation as a function of propagation loss for $\alpha_{max} = 0.25, 0.1, 0.01$.

The spectral correlation function shown in Fig. 1c also indicates a potential disadvantage of the path mismatched approach. The output of a single-mode path-mismatched interferometer will exhibit periodic modulation with frequency. This effect can be seen in Fig. 1c where the correlation initially decreases, but then increases again at the free spectral range (FSR) of 1 GHz. The FSR is set by the path mismatch of $\Delta L = 10\ cm$ and calculated as $FSR = c/(n\Delta L)$. To break this periodicity, we constructed the interferometer using multimode waveguides and relied on the frequency dependence of the speckle pattern formed at the end of the multimode waveguide to break this periodicity. As we will show, the experimental spectral correlation function still exhibits a periodic structure, but the correlation is reduced with each subsequent peak, allowing the spectral reconstruction algorithm to recover the correct spectrum. Note that the simple model used to generate Fig. 1c overestimates the extent of the re-correlation using multimode waveguides, since it neglects the wavelength dependence of the optical mode profiles and wavelength-dependent scattering or coupling coefficients in the waveguides. A more sophisticated design could suppress the nearby FSR re-correlation peaks by using several path mismatched delay lines, each with coupling coefficients designed to balance the optical power through each path after accounting for attenuation. In the experiments reported below, we exploit this approach using two path-mismatched interferometers of different length to suppress the re-correlation peak. As we will show experimentally, the remaining re-correlation peak does not degrade the performance of the RF spectrum analyzer.

**Operating Principle of the PIC-based Speckle Spectrometer**

A schematic of the PIC based speckle spectrometer that we designed and fabricated is shown in Fig. 2c. Light is coupled onto the PIC in a single mode waveguide and divided into two paths with an inverse designed splitter. The first splitter directs 20% of the optical power to a "low resolution" path mismatch interferometer with a 1 cm path mismatch and 80% of the optical power to a "high resolution" path mismatch interferometer with a 10 cm path mismatch. The 10 cm path mismatch was selected for a maximum acceptable attenuation of 10 dB assuming a waveguide propagation loss of 1 dB/cm. The arms of the path-mismatch interferometer are constructed using 20 $\mu m$ wide waveguides, supporting 42 spatial modes. For the low-resolution interferometer a splitter directed 55% of the incident optical power to the 1 cm long waveguide and for the high-resolution interferometer a splitter directed 90% of the incident optical power to the 10 cm long waveguide. The splitters were inverse designed using topology optimization (see Methods for more details) and the split ratios were designed assuming 1 dB/cm waveguide loss (based on previous experiments) to ensure equal power from each of the 4 multimode waveguides. The multimode waveguides in each interferometer were recombined in a 45 $\mu m$ wide waveguide supporting 93 spatial modes. At this point, the 45 $\mu m$ wide waveguide can be directed to an array of detectors to record the speckle pattern. In this proof-of-principle demonstration, we used an InGaAs camera and a microscope to image the speckle pattern scattered from a ridge where the photodetectors would normally be placed to avoid the complexity of integrating photodetectors.

**PIC-based RF Spectrum Analyzer Design**

To use the PIC-based speckle spectrometer as an RF spectrum analyzer, we first need to encode the RF signal in the optical domain. In this work, we consider two approaches to encode the RF signal.

The first approach is based on single sideband encoding, as shown in Fig. 2(a). In this case, the amplified RF signal drives an electro-optic intensity modulator (EOM) in dual sideband, suppressed carrier mode, creating two replicas of the RF signal as side-bands on either side of the laser frequency. An optical filter is then used to select one of these sidebands, resulting in single-sideband modulation. Note that single-sideband modulation is not strictly required, but it provides higher speckle contrast than dual-sideband modulation and a more straight-forward characterization of the optical performance of the PIC based spectrometer. Here, we used a tunable optical filter with a 10 GHz passband centered 15 GHz above the laser frequency, allowing the system to detect RF signals from 10-20 GHz (matching the bandwidth of the RF signal generator available in our lab). As shown in Fig. 2d, the optical spectrum coupled into the speckle spectrometer is a replica of the original RF spectrum, upshifted by the laser frequency (~193.2 THz). As a result, the resolution of the RF spectrum analyzer is set by the resolution of the PIC-based speckle spectrometer. This configuration also allowed us to characterize the performance of the PIC-based speckle spectrometer by probing the PIC with 1 optical frequency at a time.

The second approach, shown in Fig. 2b, uses a path-mismatched RF interferometer to improve the resolution of the RF spectrum analyzer. In this case, the RF signal is directed to a power divider which splits the RF signal into two paths. One of these arms is directed to a 2 m long coaxial cable. The two arms are then connected to a pair of matched RF amplifiers and recombined on an RF 90-degree hybrid. The outputs of the 90-degree hybrid are directed to two EOMs which are both driven in dual-sideband suppressed carrier mode, encoding the RF signal on separate carriers (lasers operating at 193.2 and 194.4 THz in this work). The sidebands encoded on the two carriers are combined on a wavelength division multiplexing (WDM) filter and then coupled onto the PIC. The relative RF power directed to each EOM varies with the RF frequency depending on the RF phase difference through the path mismatched interferometer. For a 2 m delay line, the RF power will cycle between driving EOM-1 and EOM-2 every 100 MHz ($FSR_{RF} = c/(n\Delta L_{RF})$, where $n \approx 1.5$ for the coaxial cable and $\Delta L_{RF}$ is the 2 m path mismatch). This will cause the speckle pattern to oscillate as the optical spectrum incident on the PIC changes from sidebands centered around laser-1 to sidebands centered around laser-2. Of course, this oscillation will continue to repeat, but as long as the speckle spectrometer can discriminate between RF frequencies separated by 100 MHz (determined by the spectral correlation width of the PIC and Signal-to-Noise Ratio, SNR), we can break the ambiguity and use this oscillation to improve the resolution of the overall RF spectrum analyzer. Note that in this case we opt to omit the optical filter following the EOMs used in the single sideband approach for simplicity at the cost of a minor reduction in speckle contrast.

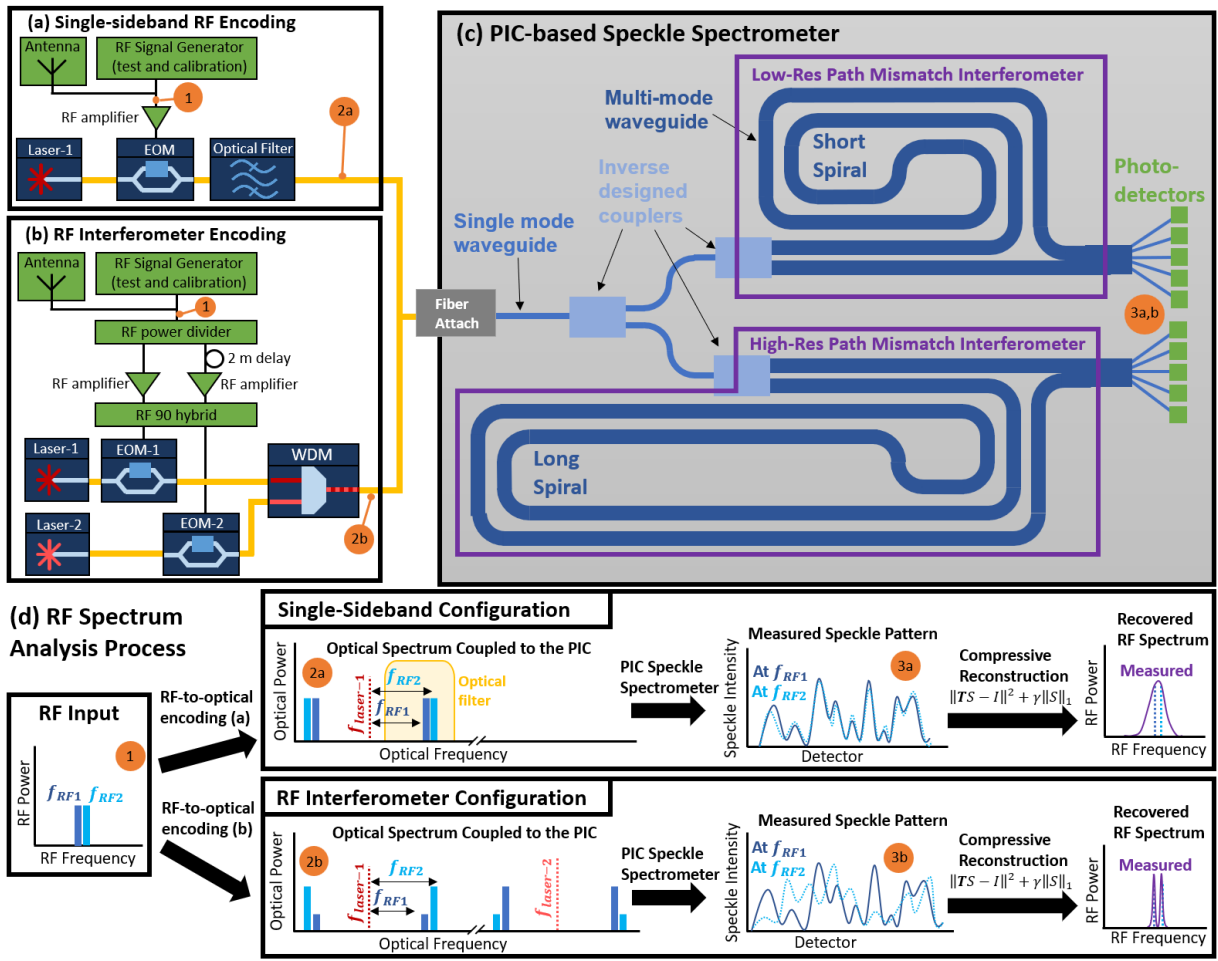

**Fig. 2. PIC-based RF spectrum analyzer architecture. a** Single-sideband RF-to-optical encoding scheme. **b** RF interferometer RF-to-optical encoding scheme. **c** PIC based speckle spectrometer layout consisting of 2 path-mismatched interferometers constructed using multimode waveguides. **d** Lower panels show how two nearby RF tones are mapped to the optical domain and then to a speckle pattern using the two RF encoding schemes. The RF interferometer encoding scheme enables higher resolution by encoding the RF signal on two optical carriers with frequency-dependent amplitude.

The panels in Fig. 2d show a simplified example of how the speckle spectrometer detects two nearby RF tones using the two encoding schemes. In the single sideband configuration, the optical spectrum coupled onto the PIC will consist of two optical tones with the same spacing and relative amplitude as the RF tones. If this spacing is less than the spectral correlation width, these tones will produce very similar speckle patterns and the speckle spectrometer will have difficulty resolving the two tones. In the RF interferometer configuration, the optical spectrum generated by the two tones is shown in the panel marked 2b. In this case, the RF power driving EOM-1 and EOM-2 is very different for the two RF tones, resulting in sidebands with different amplitude centered around the two lasers. These two optical spectra will generate very different speckle patterns, and the PIC-based RF spectrum analyzer can easily resolve these two tones.

**Fabrication and Calibration of the PIC-based Speckle Spectrometer**

The PIC-based speckle spectrometer design shown in Fig. 2c was fabricated on a 250 nm silicon-on-insulator platform. Details of the fabrication process are given in Methods. A microscope image of the fabricated device is shown in Fig. 3c, along with a scanning electron micrograph (SEM) and simulated electric field profile of one of the fabricated inverse designed splitters in Fig. 3a,b. In addition to the split ratios, the splitters were optimized via shape optimization for broadband operation (<2.5% variation in the split ratio over 300 nm, details in Methods). In this work, we used a polarization-maintaining lensed fiber to couple light onto the chip, rather than the permanent fiber-attach envisioned in the packaged system shown in Fig. 2. In addition, since the passive device did not include integrated photodetectors, we etched "detector ridges" where the detector array would have been placed. The out-of-plane scattered light from the ridges is proportional to the in-plane intensity of the propagating light[39] which is then imaged using an InGaAs camera and a microscope with a 50x microscope objective. A typical image recorded on the camera is shown in Fig. 3d. The speckle patterns recorded at the end of the "low resolution" and "high resolution" interferometers are marked in the image.

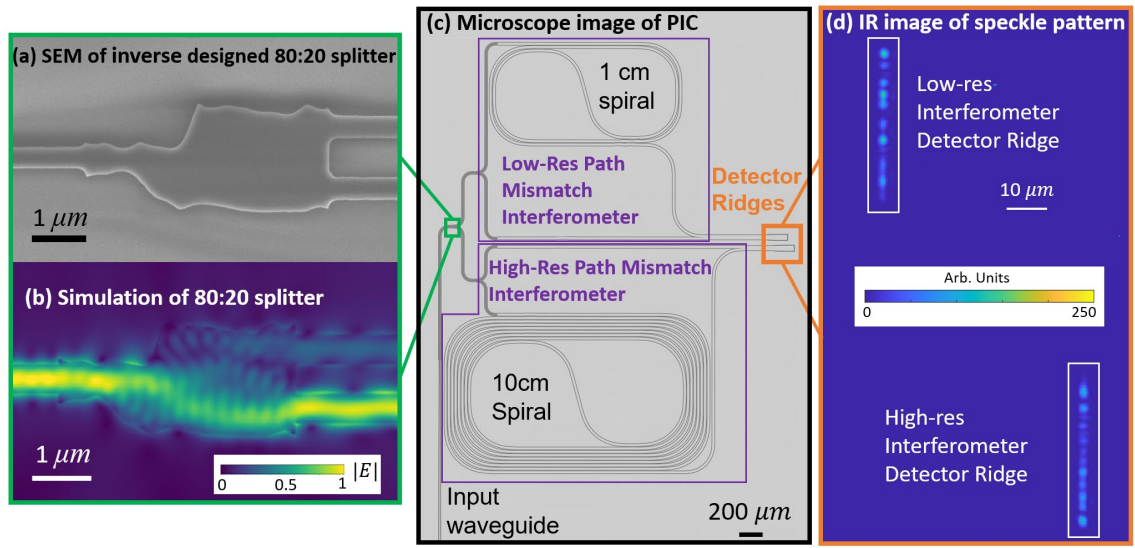

Fig. 3. Fabricated PIC-based speckle spectrometer. **a** Scanning electron microscope (SEM) image of inverse designed broadband 80:20 splitter. **b** Simulation of electric field amplitude in 80:20 splitter. **c** Microscope image of the fabricated PIC-based speckle spectrometer. **b** Image of the detector ridges recorded on the infrared (IR) camera showing the speckle patterns formed at the end of the two interferometers.

Before using the PIC as a speckle spectrometer, we need to calibrate the device by measuring the transmission matrix, $T$, that describes the speckle pattern produced by each RF frequency of interest[34]. The transmission matrix is an $M \times N$ matrix that converts the input RF spectrum, $S$, consisting of $N$ spectral channels (given by the desired bandwidth divided by the spectral resolution), to the speckle pattern $I$, consisting of $M$ spatial samples (pixels in this case, or detectors in the future). To calibrate the system, we record the speckle pattern, $I$, formed at each RF frequency in $S$ to construct the transmission matrix one column at a time. After calibration, we use a compressive sensing approach to recover the input spectrum, $S$, from the measured speckle pattern, $I$, by solving the minimization problem: $S = \underset{S}{\mathrm{argmin}}[\ \|T S - I\|_2^2 + \gamma \|S\|_1\ ]$, where $\gamma$ is the sparsity parameter[31,40].

**Single-Sideband RF Encoding Experimental Results**

We first characterized the performance of the PIC using the single-sideband based RF encoding scheme shown in Fig. 2a. Using an RF signal generator, we calibrated the transmission matrix at RF frequencies from 10 to 20 GHz in 50 MHz steps. The measured transmission matrices for the high-resolution and low-resolution interferometers on the PIC are shown in Fig. 4a,b. The transmission matrix recorded using the high-resolution interferometer shows a periodic modulation due to the FSR of the interferometer, overlaid with the gradual decorrelation due to the multimode waveguides. The spectral correlation function (see definition in Ref. [34]) shown in Fig. 4c reveals that the high-resolution interferometer rapidly decorrelates, reaching 50% correlation for a frequency shift of ~150 MHz. The correlation then increases and peaks at the FSR of ~700 MHz with a correlation of ~80% for the first re-correlation peak. The low-resolution interferometer produces a similar transmission matrix, but with a ~10x larger initial 50% decorrelation point of ~1 GHz and 10x larger FSR of ~7 GHz due to the 10x shorter path-mismatch. In this work, we concatenated the high-resolution and low-resolution detectors to form a single combined transmission matrix. The correlation function of the combined transmission matrix is also presented in Fig. 4c, showing how an additional pathlength can partially suppress the re-correlation peaks. The first re-correlation peak for the combined transmission matrix has a correlation of 72%, sufficiently low for the speckle spectrometer to discriminate between RF inputs separated by 700 MHz.

To test the RF spectrum analyzer, we recorded the transmission matrix twice and used the first measurement for calibration and the second to test the sensor. Since the speckle patterns produced by different RF spectral channels add linearly and incoherently at detection, we can use the second transmission matrix to synthesize test speckle patterns that would be produced by an arbitrary RF spectrum, $S_{test}$, as $I_{test} = T_{test} S_{test}$, where $T_{test}$ is the second measurement of the transmission matrix. Using a compressive sensing lasso algorithm[41] (see Methods for details) we can then recover the input spectrum (we used a sparsity factor of $\gamma = 1$). First, we tested the RF spectrum analyzer in 50 MHz steps across the 10 GHz operating bandwidth. As shown in Fig. 5a, the spectrum analyzer successfully recovers the frequency of individual tones in 50 MHz steps across the 10 GHz bandwidth. As shown in Fig. 5b,c, the spectrum analyzer can also resolve two tones separated by 100 MHz, as well as more complex spectra consisting of multiple tones with varying amplitudes. These examples confirmed that the re-correlation peak in the spectral correlation function did not degrade the speckle spectrometer performance.

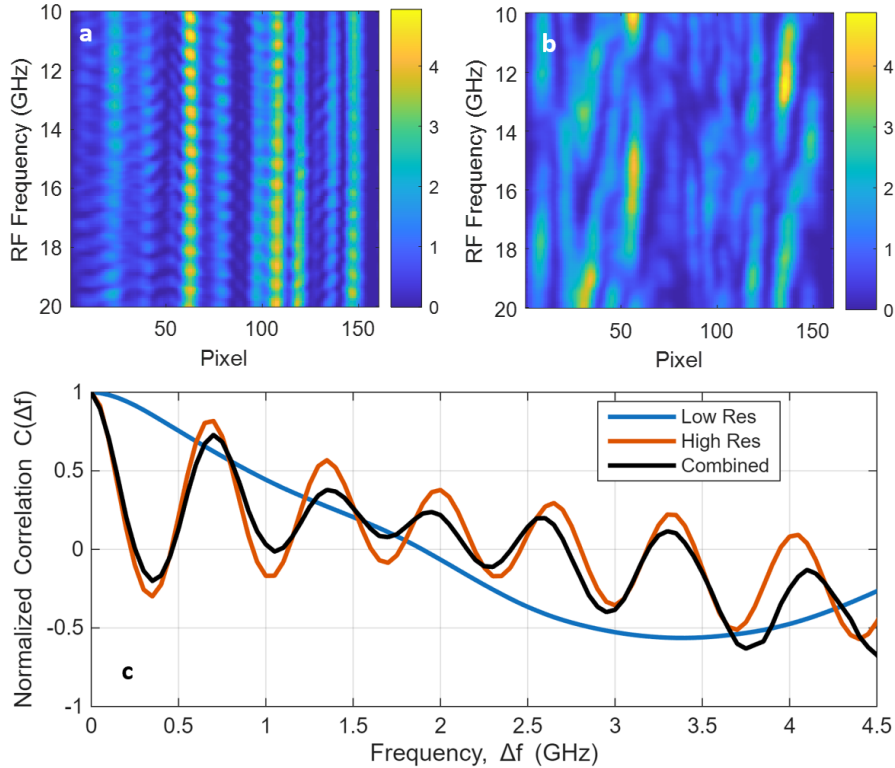

**Fig. 4. Transmission matrix using single sideband RF encoding. a** Transmission matrix recorded at the end of the high-resolution (10 cm path mismatch) interferometer using single-sideband RF-to-optical encoding. **b** Transmission matrix at the end of the low-resolution (1 cm path mismatch) interferometer. **c** Spectral correlation function for both transmission matrices and the combined transmission matrix. The high-resolution interferometer decorrelates by 50% after 150 MHz, compared to ~1 GHz for the low-resolution interferometer.

The 100 MHz spectral resolution of the PIC corresponds to just 0.8 pm at 1550 nm, and represents the highest resolution on-chip spectrometer reported to date, to the best of our knowledge (see Refs. [32,36] for recent summaries of the resolution reported for on-chip spectrometers). This resolution is possible due to the inverse designed splitters which allowed us to compensate for waveguide loss. Further improving the spectral resolution of the PIC-based speckle spectrometer would require a longer path mismatch, which would introduce more loss. To some extent, this could be compensated for using higher input power and adjusting the splitter design to direct even more power into the longer waveguide. However, to achieve 10x higher resolution, we would need a 100 cm long waveguide which would introduce unacceptably high loss of 100 dB using our current fabrication process. The RF interferometer based encoding scheme shown in Fig. 2b provides an alternative path to improve the resolution of the overall RF spectrum analyzer, as we will show in the next section.

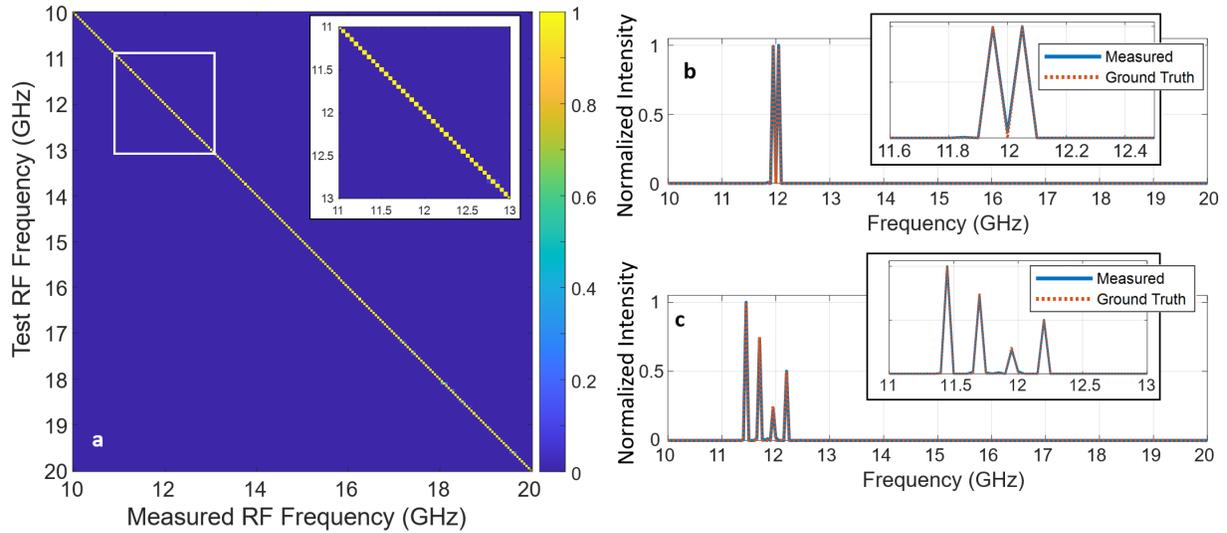

**Fig. 5. Measured RF spectra using single-sideband RF encoding. a** Recovered single RF tones in 50 MHz steps across the 10 GHz operating bandwidth. **b** Recovered RF spectrum consisting of two tones separated by 100 MHz near 12 GHz. **c** Recovered RF spectrum consisting of 4 tones with varying amplitude.

**RF Interferometer Encoding Experimental Results**

We constructed the RF interferometer encoding scheme shown in Fig. 2b using continuous wave (CW) seed lasers at 193.2 THz and 194.4 THz (these lasers were available in our lab; the precise wavelengths are not crucial as long as they are sufficiently separated to prevent RF sidebands encoded on the two lasers from overlapping). We used a fiber-coupled wavelength division multiplexing (WDM) filter to combine the two signals before coupling them onto the PIC.

Aside from these changes to the RF encoding scheme, the overall RF spectrum analyzer was calibrated and tested using the same approach described above. The RF signal generator was first used to calibrate the system by recording the speckle pattern produced by RF signals from 10 to 20 GHz. In this case, we measured the speckle patterns in steps of 5 MHz. Magnified views of the transmission matrices from 10 to 11 GHz are shown in Fig. 6a,b. In this case, both transmission matrices show periodic oscillations due to the RF interferometer. The transmission matrix recorded using the "high-resolution" interferometer also shows an oscillation with a ~700 MHz FSR from the optical interferometer. The spectral correlation patterns presented in Fig. 6c show that the RF interferometer introduces an oscillation with the expected 100 MHz period based on the 2 m delay. The correlation function for the combined transmission matrix rapidly de-correlates, reaching 50% decorrelation after 25 MHz. Although the first re-correlation peak at 100 MHz is higher than in Fig.4c (since this is based on a single-mode RF interferometer), we will show that the speckle spectrometer can still discriminate between RF tones separated by 100 MHz.

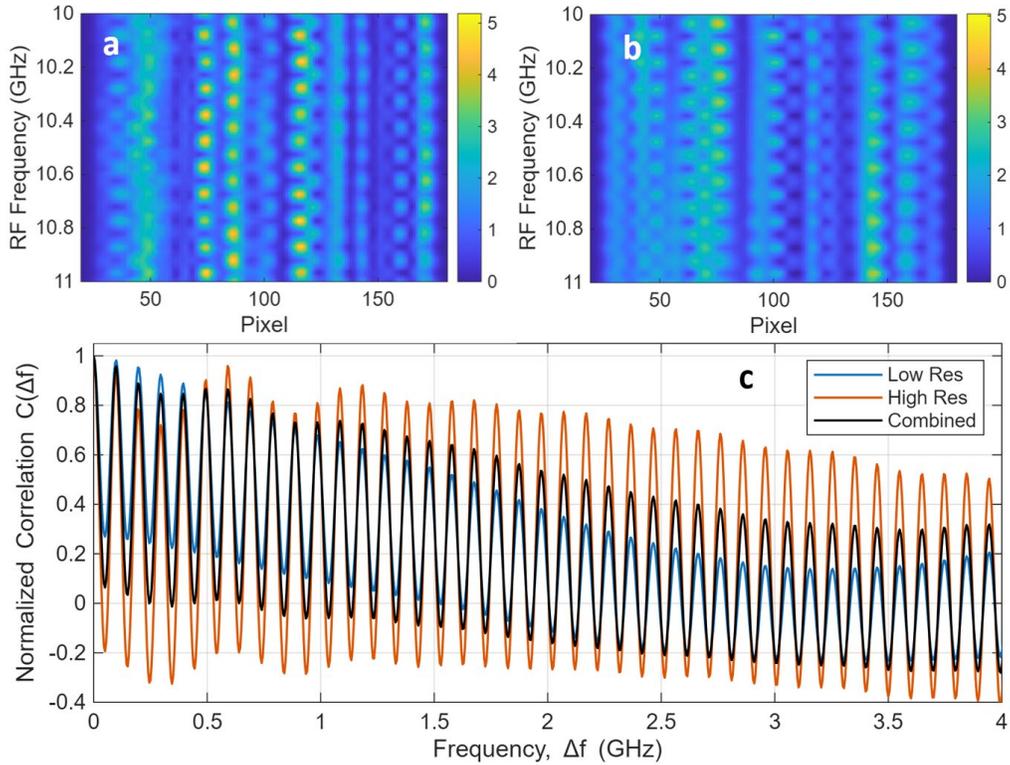

**Fig. 6. Transmission matrix using RF interferometer encoding. a** Transmission matrix recorded at the end of the high-resolution (10 cm path mismatch) interferometer using the RF-interferometer encoding. **b** Transmission matrix at the end of the low-resolution (1 cm path mismatch) interferometer. **c** Spectral correlation function for the low resolution, high resolution, and combined transmission matrices.

We then repeated the same tests reported above, recording 1 transmission matrix to use for spectral recovery and a second for testing. In this case, the transmission matrix included 2000 spectral channels (5 MHz steps from 10 GHz to 20 GHz). As shown in Fig. 7(a), the RF spectrum analyzer was able to measure individual tones across the entire 10 GHz bandwidth in steps of 5 MHz. The RF interferometer encoding scheme also enabled the RF spectrum analyzer to resolve two lines separated by only 10 MHz (Fig. 7b), and reconstruct more complex spectra consisting of multiple lines with varying amplitude (Fig. 7c). Achieving this level of resolution without the RF interferometer would have required an optical spectrometer with 10 MHz resolution (0.08 pm at 1550 nm). Based on the modeling presented in Fig. 1b, this is theoretically possible, but would likely require an improved fabrication processes with propagation loss below 0.1 dB/cm. Note that if reduced propagation loss enabled an optical spectrometer with higher native resolution, the RF interferometer encoding scheme could still be used to provide a ~10x improvement in the RF resolution beyond the resolution of the optical spectrometer.

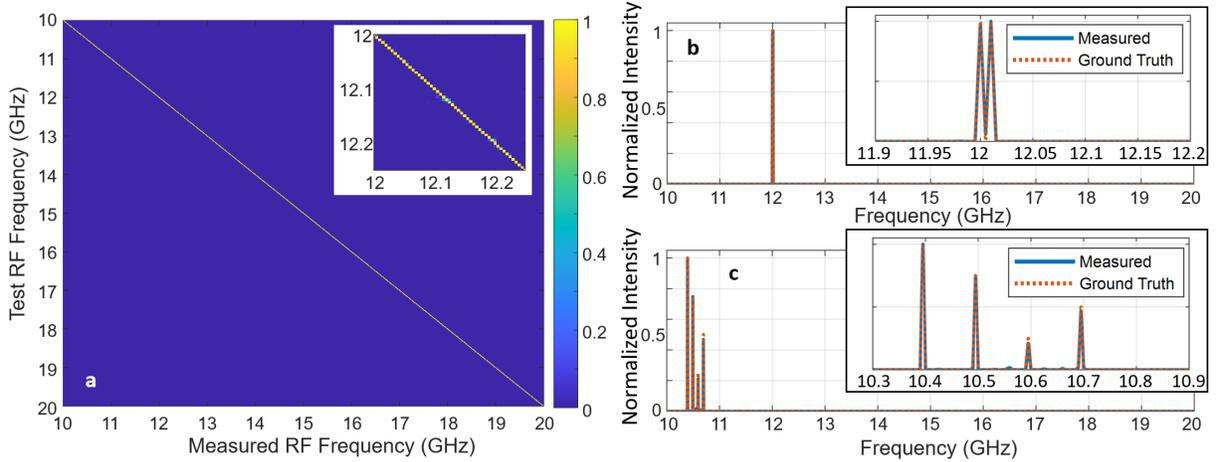

**Fig. 7. Measured RF spectra using the RF-interferometer based encoding scheme. a** Recovered single-frequency RF tones in 5 MHz steps across the 10 GHz operating bandwidth. **b** Recovered RF spectrum consisting of two tones separated by 10 MHz near 12 GHz. **c** Recovered RF spectrum consisting of 4 tones with varying amplitudes.

## Discussion

In this work, we presented an RF spectrum analyzer that combined a novel PIC-based speckle spectrometer with a unique RF-to-optical encoding scheme. The PIC-based speckle spectrometer used path-mismatched interferometers with inverse designed splitters to compensate for waveguide loss, enabling an optical resolution of 100 MHz—the highest resolution reported for an on-chip spectrometer, to the best of our knowledge. To further improve the RF resolution, we combined this speckle spectrometer with an RF interferometer based encoding scheme which allowed the RF spectrum analyzer to reach a resolution of 10 MHz while maintaining a bandwidth of 10 GHz. While this proof-of-principle demonstration focused on RF spectrum analysis, this work also shows a promising path to CMOS compatible, low SWAP-C, high-resolution and large bandwidth spectrometer designs for a wide range of applications. To advance the maturity of this technology, there are several areas that need further improvement:

(1) **Integration**: Integrating photodetectors on the PIC and using a stable fiber-attachment would enable a practical, low-SWAP system by removing the need for a microscope and IR camera. Further improvements to integration could be achieved by integrating the fiber-coupled encoding EOMs and WDMs on the chip while the RF interferometer could be integrated on a carrier board.

(2) **Speckle spectrometer design**: For this proof-of-concept, we used 2 path-mismatched interferometers. A future device could use more pickoffs (e.g., include pickoffs after every 1 cm along a 10 cm delay-line) to suppress the periodic oscillations in the transmission matrix. Lower loss waveguides would also enable a PIC-based speckle spectrometer with higher optical resolution.

(3) **Bandwidth:** In this work, the bandwidth was limited by the bandwidth of the RF signal generator used for calibration. The other limitation on bandwidth is the compression ratio and sparsity constraint. In this work, the number of spatial channels was ~100, while the number of spectral channels reached 2000. Increasing the spatial channel count (e.g., by using wider multimode waveguides) to reduce this compression ratio could improve the recovery of dense RF spectra.

(4) **Environmental stability**: One advantage of using the RF interferometer is that it is ~10,000 times less sensitive to the environment than a similar optical delay, since it depends on the RF phase delay through the 2 m delay line (i.e., $f_{opt}/f_{RF}$ ~10,000). In general, the PIC should be temperature stabilized, although in these experiments, which were conducted over 10s of minutes, no temperature stabilization was required. Following the analysis in Ref. [31], the temperature of the PIC should be stabilized to ~10 mK, which is feasible using available temperature controllers. A number of techniques have also been introduced to mitigate the need for recalibration, even if the temperature drifts beyond this level (e.g., calibration at a variety of temperatures or fabricating the device in a material such as silicon nitride with a lower thermo-optic coefficient).

**Methods**

**Simulation of spectral correlation of path-mismatched case:** The modeling of the spectral correlation functions for the path-mismatched case followed the same approach as the model for the cavity-based design described in the main text. The only difference is that in the path-mismatched case, the electric field reaching the detector was expressed as:

$$E_{det}(f_{opt}, m) = \sum_{s=1}^{N_{s-short}} \sqrt{\kappa_{short}} A_{s-short,m} \, e^{-(\alpha/2)L_{s-short,m}} e^{i[2\pi L_{s-short,m} f_{opt}/(c/n)]}$$
$$+ \sum_{s=1}^{N_{s-long}} \sqrt{\kappa_{long}} A_{s-long,m} \, e^{-(\alpha/2)L_{s-long,m}} e^{i[2\pi L_{s-long,m} f_{opt}/(c/n)]}$$

which includes a summation over $N_{s-short}$ "short" path-lengths and a summation over $N_{s-long}$ "long" pathlengths. The distributions, $L_{s-short,m}$ and $L_{s-long,m}$ account for the spread in pathlengths in short and long multimode waveguide. To simulate the path-mismatched interferometer scheme for the same number of scattering paths ($N_s$ = 1000), we set $N_{s-short} = 500$ and $N_{s-long} = 500$ and set the spread in pathlengths distributed around the short and long paths to be $L_{s-long}/100$.

**PIC-spectrometer fabrication:** The PIC-spectrometer was fabricated using commercially available silicon-on-insulator (SOI) wafers. The wafers consisted of 250 nm silicon on top of a 3 µm buried oxide. The path mismatched spirals and inverse designed splitters comprising the spectrometer were fabricated as a single exposure step using a positive tone ZEP resist followed by electron beam lithography and an optimized $C_4F_8$ and $SF_6$ -based inductively-coupled plasma reactive ion etching process.

**Experimental setup:** We characterized the PIC based RF spectrum analyzer using the setup shown in Fig. 2.

*RF Setup:* The RF signal was provided by an RF signal generator (Keysight MXG N5183B) which was amplified using a Minicircuits ZVA-183-S+ RF amp to ~23 dBm. The RF interferometer encoding scheme used an RF power divider (Krytar 6005180) and an RF 90-degree hybrid (Narda 43568).

*Optical setup:* The optical carriers were provided by narrow linewidth CW seed lasers operating at 193.2 and 194.4 THz with ~10 mW output power (Rio Orion). The lasers were fiber coupled to 20 GHz bandwidth electro-optic intensity modulators (EOSpace). The single-sideband RF encoding scheme used a Santec narrowband tunable filter (OTF-980) while the RF-interferometer encoding scheme used a fiber-coupled 100 GHz wide WDM filter to combine the signals from the two seed lasers. We used polarization maintaining fiber to connect each component and a PM lensed fiber was used to couple to the PIC. The speckle patterns scattered from the PIC were recorded on an InGaAs camera (Xenics Cheetah) using a 50x long working distance microscope objective.

**Inverse design of splitters:** The designs for the 45:55, 80:20, and 90:10 splitters were optimized using Lumerical's lumopt inverse design framework via shape optimization of the polygon geometry, leveraging finite-difference time-domain (FDTD) simulations. A gradient-based algorithm was employed to iteratively refine the device geometry, aiming to maximize transmission, here the device was optimized as a combiner, where light was input to each of the two 'output' ports and transmission into the 'input' port was measured. The initial polygon geometry was parameterized as a trapezoid with a linear slope connecting the input and output ports. The objective function for the optimization was defined as the transmission into the single port when the desired amplitudes were excited at the two input ports. The optimization and inverse design workflow consisted of four steps: (1) initialization of design parameters, (2) FDTD simulation to evaluate the objective function, (3) gradient calculation using the adjoint method, and (4) iterative updates to the design parameters based on gradient information. The final optimized designs were validated through additional simulations to confirm performance improvements and robustness.

**Lasso algorithm for Spectrum Recovery:** The Lasso algorithm for recovery of the input spectrum follows the approach shown in Ref. [31,40]. The intensity distribution at the output is given by $I = \boldsymbol{T}S$, where $\boldsymbol{T}$ is the transmission matrix and $S$ is the input spectrum. The input spectrum is then reconstructed by solving the minimization problem

$$S = \underset{S}{\mathrm{argmin}}[\; \|\boldsymbol{T}\,S - I\|_2^2 + \gamma \|S\|_1 \;],$$

where $\gamma$ is the sparsity parameter and is set to 1. In this problem, the ratio of the number of spatial channels (~100) to the number of spectral channels (~2000) defines the compression ratio which is as high as ~1:20. In general, a lower compression ratio will enable an even more accurate spectral reconstruction and will reduce the computational time of the Lasso algorithm for spectrum recovery (the number of spatial channels can be increased simply by using a wider multimode waveguide).

## Data availability

The source data are available from the corresponding author upon reasonable request.

**Acknowledgements**

B.R., J.B.M., M.J.M., R.T.S, and N.C acknowledge support from the U.S. Naval Research Lab (6.2 Base Program). R.S. acknowledges support by the Laboratory Directed Research and Development program at Sandia National Laboratories, a multi-mission laboratory managed and operated by National Technology and Engineering Solutions of Sandia, LLC, a wholly owned subsidiary of Honeywell International, Inc., for the U.S. Department of Energy's National Nuclear Security Administration. This work was performed in part at the Center for Integrated Nanotechnologies, an Office of Science User Facility operated for the U.S. Department of Energy Office of Science. This paper describes objective technical results and analysis. Any subjective views or opinions that might be expressed in the paper do not necessarily represent the views of the U.S. Department of Energy or the United States Government.